\documentclass[preprint, review, authoryear, 1p, 12pt]{elsarticle}

\usepackage{amssymb}
\usepackage{arydshln} 
\usepackage{rotating}
\usepackage{setspace}

\journal{Earth and Planetary Science Letters}

\newcommand\T{\rule{0pt}{2.6ex}}

\newcommand\TT{\rule{0pt}{3.5ex}}
\newcommand\BB{\rule[-2.2ex]{0pt}{0pt}}

\begin{document}

\begin{frontmatter}

\title{Comparison of the Oxidation State of Fe in Comet 81P/Wild 2 and Chondritic-Porous Interplanetary Dust Particles}

\author[label1]{R. C. Ogliore\corref{cor1}}
\cortext[cor1]{Corresponding author}
\ead{ogliore@ssl.berkeley.edu}
\author[label1]{A. L. Butterworth}
\author[label2]{S. C. Fakra}
\author[label1]{Z. Gainsforth}
\author[label2]{M. A. Marcus}
\author[label1]{and A. J. Westphal}
\address[label1]{Space Sciences Laboratory, University of California at Berkeley, Berkeley, CA 94720, USA}
\address[label2]{Advanced Light Source, Lawrence Berkeley National Laboratory, Berkeley, CA 94720, USA}

\begin{abstract}
The fragile structure of chondritic-porous interplanetary dust particles (CP- IDPs) and their minimal parent-body alteration have led researchers to believe these particles originate in comets rather than asteroids where aqueous and thermal alteration have occurred. The solar elemental abundances and atmospheric entry speed of CP-IDPs also suggest a cometary origin. With the return of the Stardust samples from Jupiter-family comet 81P/Wild 2, this hypothesis can be tested. We have measured the Fe oxidation state of 15 CP-IDPs and 194 Stardust fragments using a synchrotron-based x-ray microprobe. We analyzed $\sim$300 nanograms of Wild 2 material -- three orders of magnitude more material than other analyses comparing Wild 2 and CP-IDPs. The Fe oxidation state of these two samples of material are $>$2$\sigma$ different: the CP-IDPs are more oxidized than the Wild 2 grains. We conclude that comet Wild 2 contains material that formed at a lower oxygen fugacity than the parent body, or parent bodies, of CP-IDPs. If all Jupiter-family comets are similar, they do not appear to be consistent with the origin of CP-IDPs. However, comets that formed from a different mix of nebular material and are more oxidized than Wild 2 could be the source of CP-IDPs.
\end{abstract}


\end{frontmatter}

\section{Introduction}
\label{intro}
Chondritic-porous interplanetary dust particles (CP-IDPs), collected in the stratosphere by high-altitude aircraft, are widely thought to originate in comets \citep{Schramm:1989p3433,Bradley:1994p2672}. The collisional history of asteroids makes it unlikely that these particles (fragile, porous aggregates of small grains) are asteroidal \citep{Brownlee:1985p2339}.  The atmospheric entry speed of CP-IDPs, as deduced from thermal release profiles of solar wind He \citep{Nier:1992p3404}, is consistent with cometary, rather than asteroidal, orbits \citep{Brownlee:1995p2673}. Infrared and electron beam studies of IDPs show that those consisting of mostly anhydrous minerals are usually chondritic-porous, whereas IDPs rich in phyllosilicates are usually chondritic-smooth \citep{Sandford:1985p3456, Bradley:1986p3458}. The presence of GEMS (glass with embedded metal and sulfides) in IDPs indicates a lack of thermal alteration \citep{Bradley:1994p2672}. CP-IDPs, therefore, show very little parent-body alteration and must originate from either anhydrous objects or hydrous objects that have been kept at very low temperature -- again, consistent with cometary origin. Nevertheless, a cometary origin of CP-IDPs is not universally accepted \citep{Flynn:1992p3174, Thomas:1995p3175}. With the return of samples from comet 81P/Wild 2 by NASA's Stardust mission \citep{Brownlee:2006p1376}, it is now possible to compare CP-IDPs with material from this Jupiter-family comet. We seek to prove or disprove, with a known level of confidence, the hypothesis that CP-IDPs originate from parent bodies with the composition of comet Wild 2.

The oxidation state of Fe is a clear mineralogic indicator of the oxidation state of meteorites \citep{Rubin:1988p2665}. Meteorite groups are in fact distinguished from each other by their differing Fe oxidation states, as Urey and Craig first reported more than fifty years ago \citep{Urey:1953p1396}. Parent body processing can drastically change the Fe oxidation state of some of the meteorite groups, obscuring information about the original oxidation state the material acquired in the solar nebula \citep{Rubin:1988p2665}. Carbonates and Fe-bearing crystalline silicates, possible products of aqueous alteration on asteroids \citep{Krot:1995p3169}, are found alongside anhydrous minerals in the same Stardust track \citep{Flynn:2008p2675}, and therefore co-existed in one large aggregate in comet Wild 2 . No phyllosilicates have been unambiguously identified in the Stardust samples \citep{Zolensky:2008p1420}.  With these and other pieces of evidence, \citet{Wooden:2008p2674} argues that instead of selective aqueous alteration on submicron scales in the comet itself, grains which formed in different regions of the solar nebula under varying reduction-oxidation conditions (e.g. Mg- and Fe-rich crystalline silicates) migrated, aggregated, and formed comet Wild 2. Likewise, the oxidation state of anhydrous CP-IDPs was unaffected by parent body processing \citep{Schramm:1989p3433,Zolensky:1995p1397}, so a comparison of the Fe oxidation state of comet Wild 2 with anhydrous CP-IDPs yields insight into the nebular environment in which they formed.

We report on the oxidation state of Fe in 15 anhydrous CP-IDPs, and directly compare these measurements to the Fe oxidation state of comet Wild 2, deduced from 194 Stardust fragments in 11 aerogel tracks. 

\section{Methods}

\subsection{Sample Preparation}

\begin{figure}
        \centering
        \includegraphics[width=\textwidth]{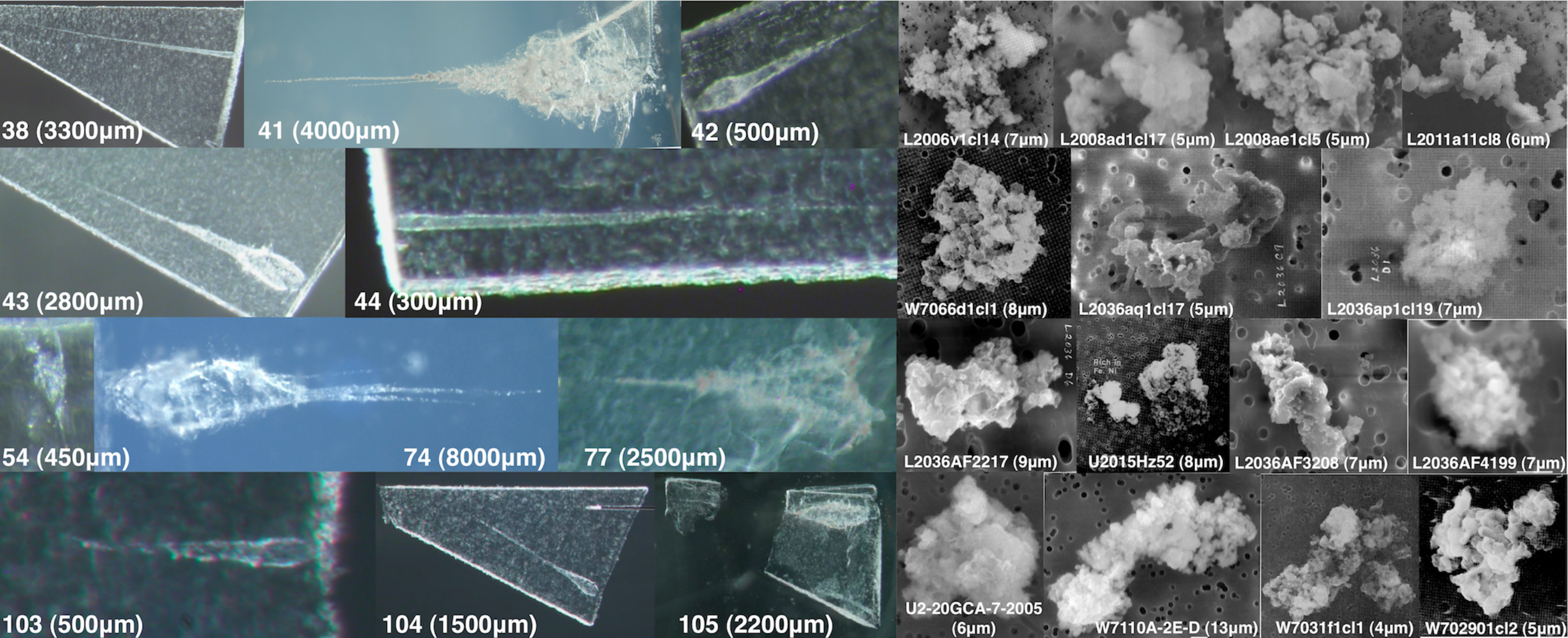}
        \caption{Optical microscope photographs of the Stardust tracks analyzed in this study (left) and SEM images of the IDPs analyzed (right). IDP SEM images are from the Cosmic Dust Catalogs except for U2-20GCA-7-2005 and W7110A-2E-D which are courtesy D. Joswiak (University of Washington).}
        \label{figure:photos}
  \end{figure}

We identified and requested a set of IDPs from volumes 16 and 17 of the Cosmic Dust Catalogs \citep{Warren:1997p2676}.  These particles were collected by NASA U-2, ER-2 and WB-57F aircraft carrying flat-plate collectors coated with silicone oil at an altitude of 20 km \citep{Warren:1994p2679}. We selected ``Cosmic"-type particles that looked like fluffy, fragile, aggregates in the SEM images given in the catalogs (see Figure \ref{figure:photos}) that also had approximately chondritic composition as shown in the catalog's energy dispersive X-ray spectra (EDX) (see Table \ref{table:elements}). We also acquired IDPs from collaborators who identified the particles as chondritic and porous. The particles were mounted at Johnson Space Center with silicone oil on a 4$\mu$m-thick ultralene film.

Cometary particles impacting the Stardust aerogel collector at 6.1 km s$^{-1}$ form tracks ranging from short and bulbous to long and narrow. We extracted 11 tracks, randomly chosen from the set of tracks nearly normal to the plane of the collector, of various morphologies and sizes (see Figure \ref{figure:photos}) in ``keystones'' (wedge-shaped pieces of aerogel) using robotically controlled micromanipulators and pulled-glass microneedles \citep{Westphal:2004p2678}. The tracks were either mounted in an ultralene sandwich or onto silicon ``microforks".

\singlespacing

\begin{sidewaystable}[ht]
\centering
\begin{tabular}{l  p{5em}  p{5em}  p{5em}  p{5em}  p{5em}  p{5em}  p{5em}}
\hline\hline  \TT \BB
IDP & Mg/Si/All. & Al/Si/All. & S/Si/All. &  Ca/Si/All. & Cr/Si/All. & Fe/Si/All. & Ni/Si/All. \\  [0.5ex]
L2006v1cl14 & 1.2 & 1.1 & 3.7 & 0.4 & 1.3 & 0.7 & 1.0 \\
L2008ad1cl17 & 1.0 & 1.0 & 1.2 & 0.6 & 0.8 & 0.8 & 1.2 \\
L2008ae1cl5 & 0.9 & 1.7 & 1.2 & 0.4 & 0.8 & 0.6 & 0.6 \\
L2011a11cl8 & 0.9 & 0.5 & 0.8 & 0.6 & 1.2 & 0.9 & 1.5 \\
L2036ap1cl19 & 1.2 & 0.7 & 2.1 & 0.3 & 1.0 & 0.8 & 0.5 \\
L2036aq1cl17 & 1.0 & 0.4 & 0.5 & 0.4 & 0.9 & 0.7 & 0.7 \\
W7031f1cl1 & 0.9 & 1.1 & 2.0 & 0.8 & 2.0 & 0.5 & 0.8 \\
W7066d1cl1 & 0.8 & 1.8 & 3.7 & 0.6 & 1.5 & 0.7 & 1.2 \\
W702901cl2 & 0.8 & 1.1 & 1.4 & 0.7 & 1.8 & 0.7 & 1.0 \\
L2036AF2217 & 1.1 & 0.8 & 2.2 & 0.6 & 1.1 & 0.8 & 0.0 \\
U2015Hz52 & 0.8 & 0.9 & 2.3 & 0.4 & 1.4 & 0.6 & 0.8 \\
L2036AF3208 & 1.0 & 1.9 & 7.2 & 0.9 & 1.8 & 0.9 & 1.0 \\
L2036Af4199 & 1.2 & 0.7 & 2.2 & 0.4 & 1.2 & 0.8 & 0.7 \\
U2-20-GCA-7-2005 & & \parbox[c]{25pc}{chondritic, as given in \citet{Dai:2002p2677}}\\
W7110-A-2E & & \parbox[c]{25pc}{chondritic, as given in \citet{Dai:2002p2677}}\\
\hline
\hline
\end{tabular}
\caption{Mg, Al, S, Ca, Cr, Fe, and Ni abundances relative to Si and Allende, as given in the Cosmic Dust Catalogs.}
\label{table:elements}
\end{sidewaystable}

\doublespacing

\subsection{X-ray microprobe analysis}

We analyzed both sets of samples on beamline 10.3.2 at the Advanced Light Source \citep{Marcus:2004p2680}.  This beamline uses a Si(111) monochromator and Kirkpatrick--Baez mirrors to focus a monochromatic x-ray beam to a spot size between 1.5 $\times$ 1.5 $\mu$m and 16 $\times$ 7 $\mu$m. We collected micro-X-ray fluorescence ($\mu$XRF) elemental maps at 10 keV incident energy for each keystone and IDP. The fluorescence signal was recorded by a seven-element Ge solid-state detector with a nominal dwell time of 100 ms per pixel. The $\mu$XRF maps allowed us to locate the IDP in the ultralene for subsequent micro-X-ray absorption near-edge structure ($\mu$XANES) spectroscopy at the Fe K-edge. For the Stardust tracks, we only performed Fe K-edge $\mu$XANES on the most Fe-rich particles that were identified by $\mu$XRF mapping. For a detailed description of the Fe XANES technique on Stardust samples, see \citet{Westphal:2009p1874}.

\subsection{Quantification of Fe oxidation state}

\begin{figure}
        \centering
        \includegraphics[width=\textwidth]{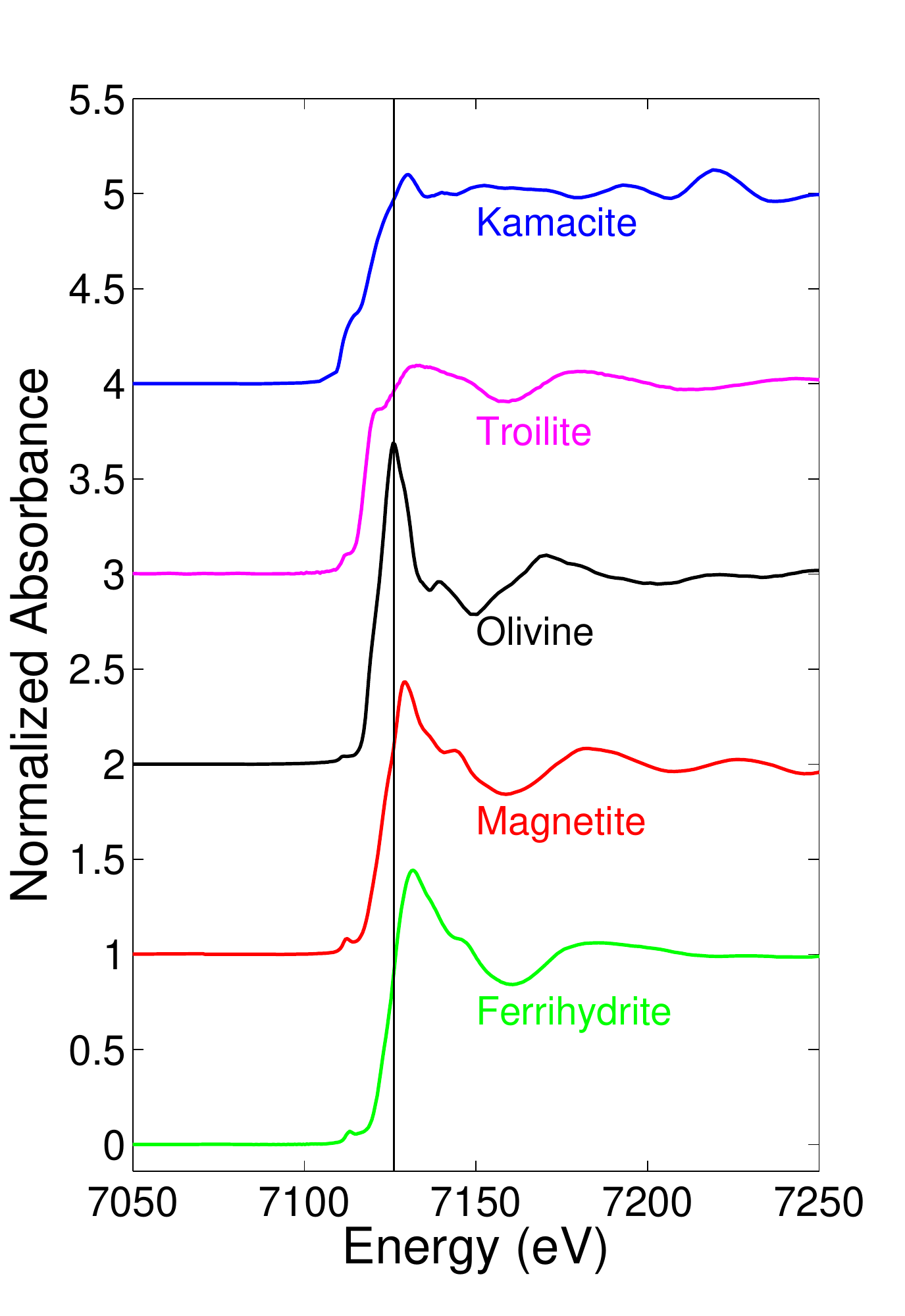}
        \caption{Fe K-edge XANES spectra of kamacite (Fe$^{0}$), troilite (sulfide), olivine (Fe$^{2+}$), magnetite (mixed Fe$^{2+}$ and Fe$^{3+}$), and ferrihydrite (Fe$^{3+}$). Ordinate is shifted for clarity, vertical line marks the main peak for olivine. Spectra were acquired to $\sim$7400 eV.}
        \label{figure:xanesstandards}
  \end{figure}

\begin{figure}
        \centering
        \includegraphics[width=\textwidth]{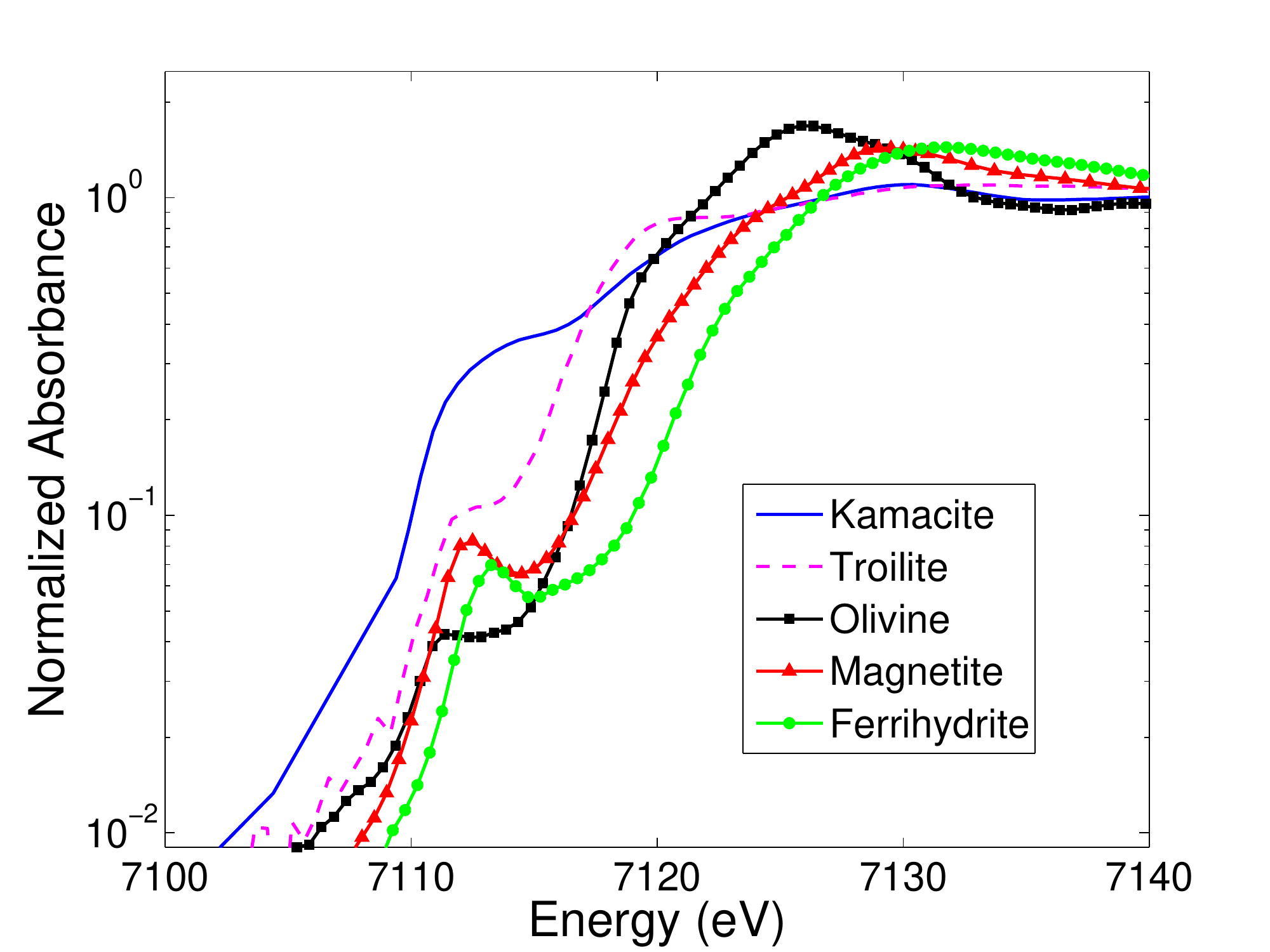}
        \caption{A closer view (with a logarithmic ordinate) of the absorption edge of the mineral standards shown in Figure \ref{figure:xanesstandards}, showing the Fe$^{0}$ standard, kamacite, absorbing first, followed by the sulfide (troilite), then Fe$^{2+}$ (olivine) and magnetite slightly later (mixed Fe$^{2+}$ and Fe$^{3+}$), and finally the most oxidized mineral, ferrihydrite (Fe$^{3+}$).}
         \label{figure:xanesstandardsedges}
  \end{figure}

The XANES spectra of multivalence elements can be used to determine oxidation state. Atoms in a reduced environment start to strongly absorb photons several eV below the energy at which the same atoms in an oxidized environment do. The energy resolution of the beamline near the Fe K-edge is $\sim$1 eV, so this shift is clearly resolved in the XANES spectra (see Figures \ref{figure:xanesstandards} and  \ref{figure:xanesstandardsedges}). By this means, the oxidation state of the analyzed material can be determined with high certainty \citep{Marcus:2008p3170}.

Fe K-edge XANES contains information about the local structure and electronic states around Fe atoms. The spectra vary between minerals and significantly between mineral groups \citep{ODay:2004p2774}. Consequently, Fe K-edge XANES allows us to distinguish between minerals of different groups, though degeneracies exist between minerals of the same group (see Figure 4 in \citet{Westphal:2009p1874}).

We acquired Fe K-edge XANES spectra of 54 minerals from 24 different mineral groups (metal, sulfide, olivine, silicate glass, carbide, orthopyroxene, clinopyroxene, phyllosilicate, and others as given in Table 1 of  \citet{Westphal:2009p1874}). The spectrum of each of these mineral standards was calibrated, pre-edge background-subtracted and post-edge normalized using standard procedures in XANES analysis \citep{Kelly:2008p2912}.  To account for x-ray dichroism in minerals of non-cubic symmetry, we acquired XANES spectra of several individual grains of our reference minerals \citep{Marcus:2008p3170,Westphal:2009p1874}. 

For most of the Stardust tracks, we acquired spectra of all fragments in the track with sufficient Fe content for XANES analysis. For the largest Stardust tracks (41, 74, and 77) we analyzed all the fragments in the terminal end of the track and a random selection of fragments in the track bulb, closer to the aerogel surface. This could introduce a selection bias if the composition of the track varies as a function of depth in the aerogel. Additionally, the x-ray beam spot size could be smaller than the particle for the largest particles we analyzed, resulting in incomplete XANES sampling of some of the large terminal particles. These selection biases were accounted for by comparing the total fluorescence intensity from the background-subtracted XRF maps of each track to the sum of the XANES edge jumps in that track (for particles optically-thin and smaller than the beam spot size, the magnitude of the edge jump is proportional to the total amount of Fe in the particle). This yields an undersampling correction factor that can be applied to the XANES measurements to properly account for the amount of Fe in the bulb and terminal fragments. Additionally, from XRF and the absorption edge jump at the Fe K-edge, we compute the fraction of Fe in the 11 Stardust tracks that we analyzed by XANES--about 20\%.

We analyzed 194 Stardust fragments in 11 tracks. The fragments varied in size from a few microns (fragments in the bulb of the track) to a few tens of microns (terminal particles of large tracks). Four of the largest terminal fragments were corrected for overabsorption (using a simple model described in \citet{Manceau:2002p2964}) because they were thick compared to the absorption depth of the incident x-ray beam. We analyzed the data using two independent methods. In the first approach, we fit the 194 Stardust spectra to our library of 56 mineral reference spectra using the non-negative least-squares fitting function \texttt{lsqnonneg} in MATLAB. In the second approach, we least-squares fit the Stardust spectra using the best linear combination fit of at most three reference spectra. The two methods produced consistent results. We report results from the first approach.

\begin{figure}
        \centering
        \includegraphics[width=\textwidth]{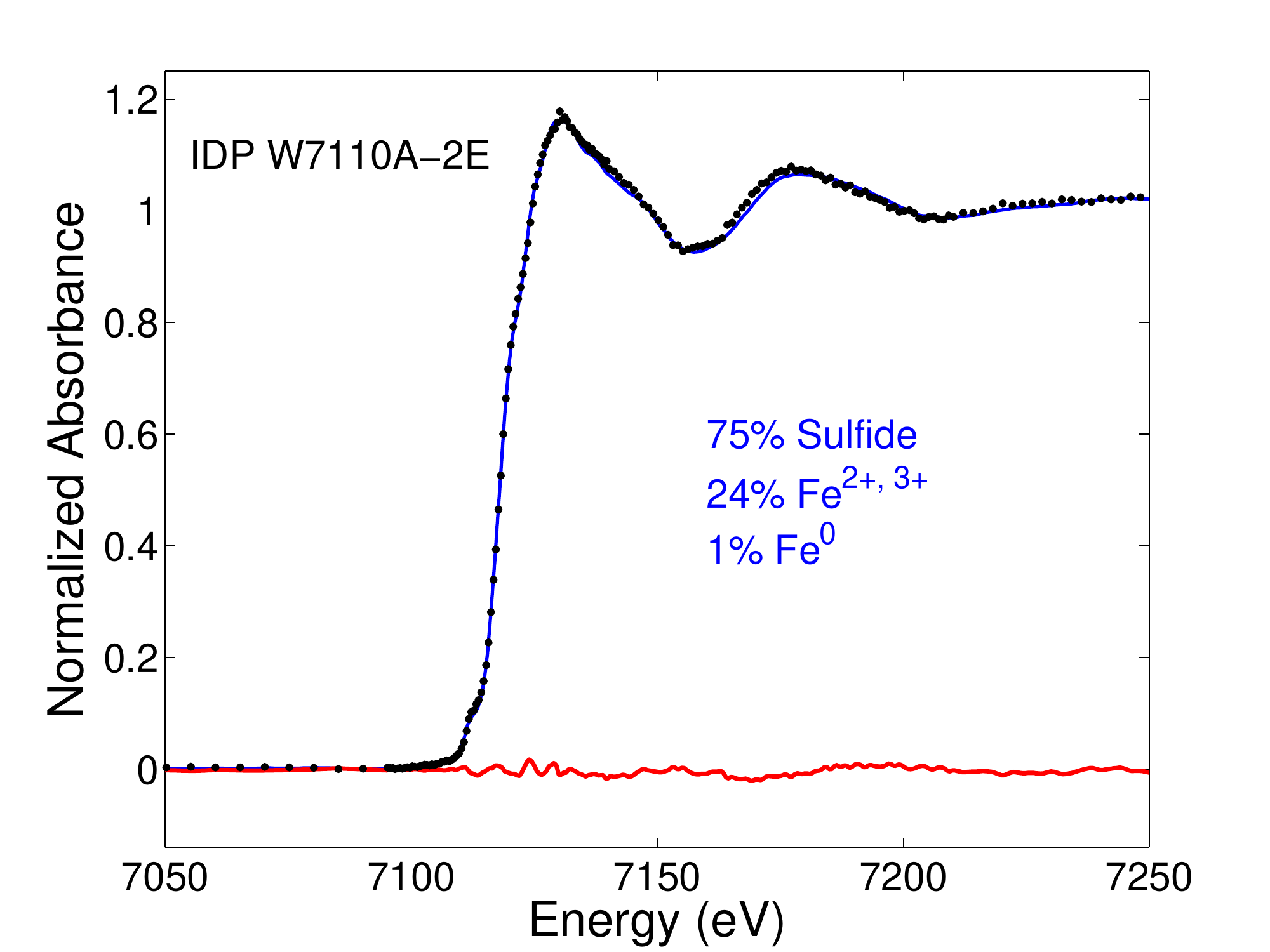}
        \caption{Fe K-edge XANES spectra for IDP W7110A-2E (black dots), its least-squares fit to mineral standards (blue), and residual of the fit (red). The particle fit to predominantly sulfide. The log of the sum of the residual of the least-squares fit for this particle is -4.3, the mean summed log residual for the 15 CP-IDPs is -4.4.}
           \label{figure:xanesfit}
  \end{figure}

We also analyzed the 15 CP-IDPs on beamline 10.3.2, using similar experimental conditions in order to directly compare the CP-IDPs with the Stardust samples.  First we located these particles in an optical microscope and encircled them with Ag paint on the ultralene to facilitate identification at the synchrotron. Once the particle was located by XRF mapping at beamline 10.3.2, we collected $\mu$XRF spectra with an incident beam energy of 10 keV and qualitatively compared this with the JSC EDX data to ensure we had located the IDP. Then we acquired Fe K-edge XANES spectra and analyzed the data using the same methods we used to analyze the Stardust tracks. An example of a Fe K-edge XANES spectrum of a CP-IDP and its fit to mineral standards is shown in Figure \ref{figure:xanesfit}.

We categorized each mineral standard as either Fe$^{0}$ (metals, silicides, and carbides), sulfide, or oxidized Fe (silicates, Fe$^{2+}$, and Fe$^{3+}$), so each CP-IDP and Wild 2 fragment was described by its fraction of Fe in these three states. The Fe oxidation state composition of the particles for each Stardust track and each CP-IDP is given in Table \ref{table:fefrac}. We weighted each particle by the magnitude of its XANES edge jump and determined the bulk Fe oxidation state composition of our samples of comet Wild 2 and CP-IDPs. 

\subsection{Determination of confidence limits}
\label{confidencelimits}

We calculated confidence limits by using the bootstrap method \citep{RChernick:1999p3002}, a Monte Carlo based approach that is appropriate to use when the dispersion and mean of the data are not known \emph{a priori}. This method is equivalent to asking: How different would the result be if we looked at a different set of 15 CP-IDPs (allowing repeats) chosen from the same ensemble? We assume that the diversity in the Fe oxidation state of CP-IDPs and comet Wild 2 is reflected in the diversity of our analyzed samples. After calculating the mean contribution of Fe$^0$, Fe sulfides, and oxidized Fe from 10$^5$ randomly chosen sets of 15 CP-IDPs, we took as our 2$\sigma$ allowed region the convex hull in the space defined by the Fe oxidation state that contained 95.45\% of the 10$^5$ numerical trials. Similarly, we calculated this region for the cometary particles taking as our set of measurements the average oxidation state in each of the 11 Stardust tracks. Using the same method, we calculated the confidence interval using the 194 Stardust fragments independently, but this resulted in smaller uncertainties (suggesting that the bulk Fe oxidation state varies systematically from one track to another). We chose to use the track-based approach which yields larger uncertainties. 

The comet Wild 2 material is 24\%$^{+6\%}_{-14\%}$ Fe$^{0}$, 47\%$^{+7\%}_{-17\%}$ Fe sulfide, and 29\%$^{+26\%}_{-11\%}$ oxidized Fe; the CP-IDPs are 1\%$^{+4\%}_{-1\%}$ Fe$^{0}$, 55\%$^{+25\%}_{-33\%}$ Fe sulfide, and 44\%$^{+31\%}_{-25\%}$ oxidized Fe (2$\sigma$ uncertainties).

\section{Results}

\begin{figure}
        \centering
        \includegraphics[width=\textwidth]{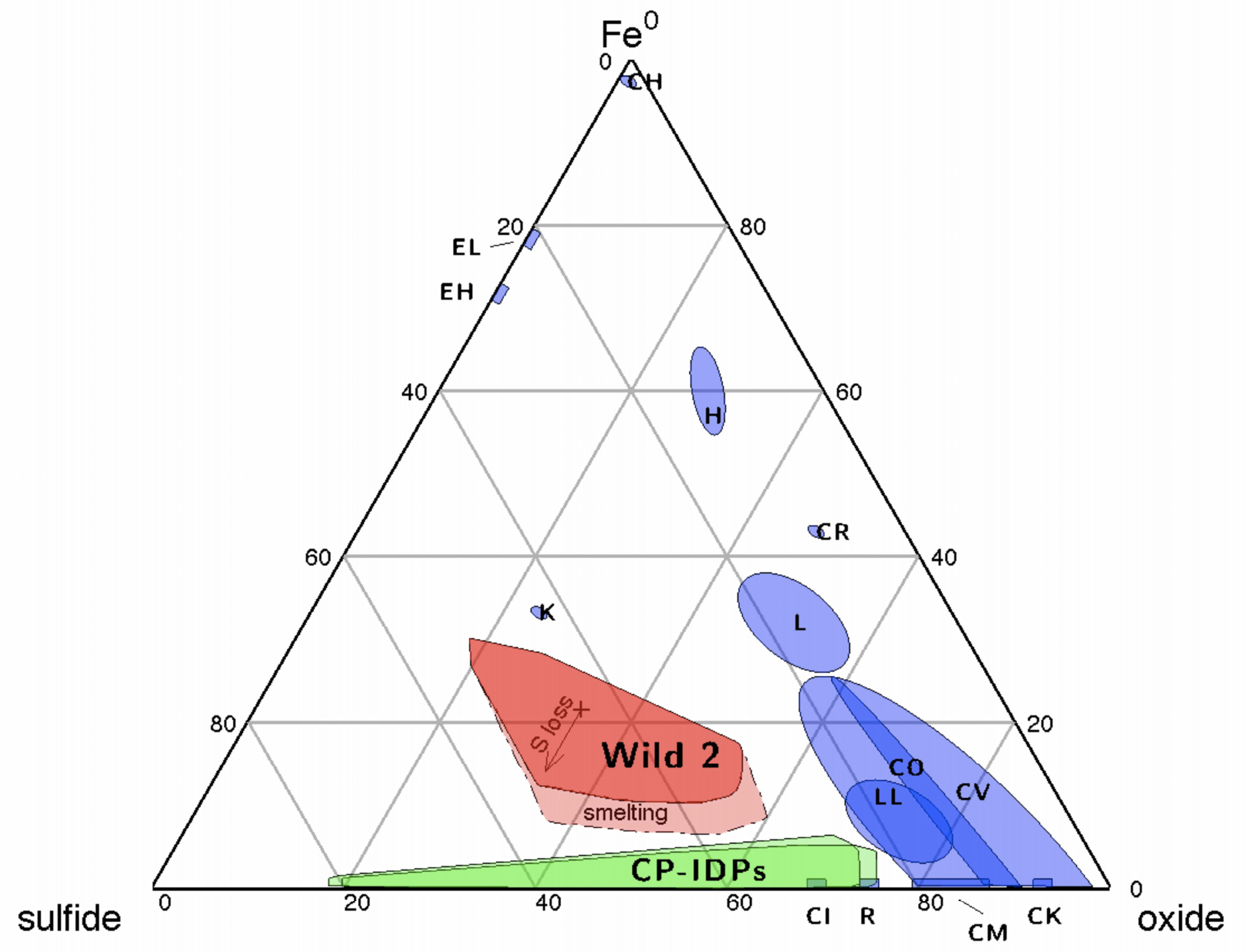}
        \caption{Ternary plot of fraction of Fe in Fe$^0$, Fe sulfides, and oxidized Fe (silicates, Fe$^{2+}$, and Fe$^{3+}$) for carbonaceous (CI, CM, CO, CK, CV, CV), ordinary (H, L, LL) and enstatite (EH, EL) chondrite groups, comet Wild 2, and CP-IDPs. Regions representing meteorite groups encompass measured dispersions within each group (except for EH, K, CR, and CI which are single measurements), whereas the CP-IDP and Wild 2 regions are 2$\sigma$ allowed regions. Magnetite in CP-IDPs originated as either oxidized Fe or a mix of sulfides and Fe$^{0}$, as described in the text, shifting the CP-IDP region to the left and up slightly in the ternary plot. Capture effects for Stardust grains are shown by the ``smelting" region (assuming all iron carbide is a reduction product that originated as oxidized Fe) and the ``S loss" arrow (reduction of sulfides to metal), an effect that is expected to be small, as described in the text.}
\label{figure:ternary}
  \end{figure}

The 2$\sigma$ allowed regions for the molar fraction of Fe in Fe$^{0}$, Fe sulfide, and oxidized Fe in comet Wild 2 and CP-IDPs are shown in Figure \ref{figure:ternary}. For comparison, the distribution of Fe in meteorites is also given. Data is from \citet{Jarosewich:1990p3052}, \citet{Yanai:1992p3053}, \citet{Mason:1962p3075}, \citet{Mason:1962p3076}, and \citet{Mason:1966p3077}. Here the regions are not 2$\sigma$ uncertainties, but rather enclose the published data.

\subsection{Effects of capture}

The presence of magnetite in IDPs can be due to heating upon entry into the atmosphere.  Electron microscopy of IDPs by \citet{Germani:1990p710} shows magnetite sometimes present as a polycrystalline decoration on sulfide grains or as a $\sim$100 nm rim, similar to a fusion crust, around the IDP (also see \citet{Zolensky:1995p1397} and references therein). Possibly, some magnetite can be indigenous to the IDP before atmospheric entry \citep{Fraundorf:1981p925}. Magnetite can be distinguished with reasonable certainty from other Fe-bearing phases in our Fe XANES library because of its mixed oxidation state (see Figure \ref{figure:xanesstandardsedges}). Table \ref{table:fefrac} shows that 4 of the 15 CP-IDPs contain at least 1\% magnetite; only 3 of the 194 Stardust fragments analyzed contain at least 1\% magnetite. We cannot rule out the possibility that the observed magnetite in the 4 CP-IDPs is due to oxidation of Fe$^{0}$ during atmospheric entry, so we assume that the magnetite originated as Fe$^{0}$ or iron-sulfide in the proportion present in the other 11 CP-IDPs (Fe$^{0}$/sulfide= 0.04). With this assumption, the 2$\sigma$ allowed region for CP-IDPs shifts slightly towards more sulfide and Fe$^{0}$, as shown in Figure \ref{figure:ternary}. A $\sim$100 nm rim of magnetite around a 7 $\mu$m IDP is only a few percent of the total volume of the entire particle, so it is possible that some of our IDPs may have a thin magnetite rim that was not well-identified by our Fe XANES measurements. \citet{MacKinnon:1987p2660} show that magnetite is observed in about half of CP-IDPs; we conclude that if a thin magnetite rim was not detected by our measurements, the amount of total Fe mischaracterized would be small and would not significantly effect our results.  

The hypervelocity capture process of cometary grains into aerogel could alter the material's oxidation state, so this effect must be accounted for. The impactor slows from 6.1 km s$^{-1}$ to rest in tens to hundreds of nanoseconds, resulting in a wide range of thermal modification of individual fragments \citep{Zolensky:2006p1378}. We have observed, using TEM and Fe XANES, reduction of Fe in basalt glass projectiles shot into aerogel at 6.1 km s$^{-1}$ \citep{Marcus:2008p45}. The Fe is reduced by a reaction with C impurities in the aerogel (a process similar to smelting) and forms iron carbide, not metal. TEM observations of small, thermally modified Stardust grains (where the smelting effect would likely be greatest), do not show significant amounts of iron carbide \citep{Leroux:2008p1416,Stodolna:2009p3426,Leroux:2009p3050}. We conclude that the iron carbide contribution in our XANES fits is probably cometary Fe$^{0}$, not smelted Fe$^{2+}$ or Fe$^{3+}$. Nonetheless, the possibility of smelting is accounted for in our uncertainties in the 2$\sigma$ allowed region shown for Wild 2 in Figure \ref{figure:ternary}. 

Because of their lower melting point, sulfides can melt during capture and become widely dispersed throughout the track after they recondense into small fragments. In this study, we collect Fe XANES from only the particles with the strongest Fe signals in our $\mu$XRF survey, so these small sulfides may be missed. Because we analyzed the most Fe-rich particles, this effect is likely to be small and would result in Wild 2 more enriched in sulfides than it appears in Figure \ref{figure:ternary}.

Sulfur can be lost from Fe-Ni sulfides in Stardust samples due to the capture process \citep{Zolensky:2006p1378}, and sulfides can be reduced to Fe metal in the form of Fe-Ni-S nanophases \citep{Ishii:2008p962} (an Fe metal core surrounded by a sulfide mantle). \citet{Leroux:2009p3050} examined two thermally modified particles by TEM, extracted from the wall of the bulbous part of Track 35, and observed the Fe-Ni-S nanophases. They hypothesize that these phases originated as FeO and were reduced during capture due to the combustion of organics and the recondensation of gaseous S$_2$ onto metal droplets. \citet{Leroux:2009p3050} conclude that the capture process tends to lower the Fe oxidation state for thermally modified particles but ``terminal particles appear relatively preserved and may exhibit their pristine redox state". 

\begin{figure}
        \centering
        \includegraphics[width=\textwidth]{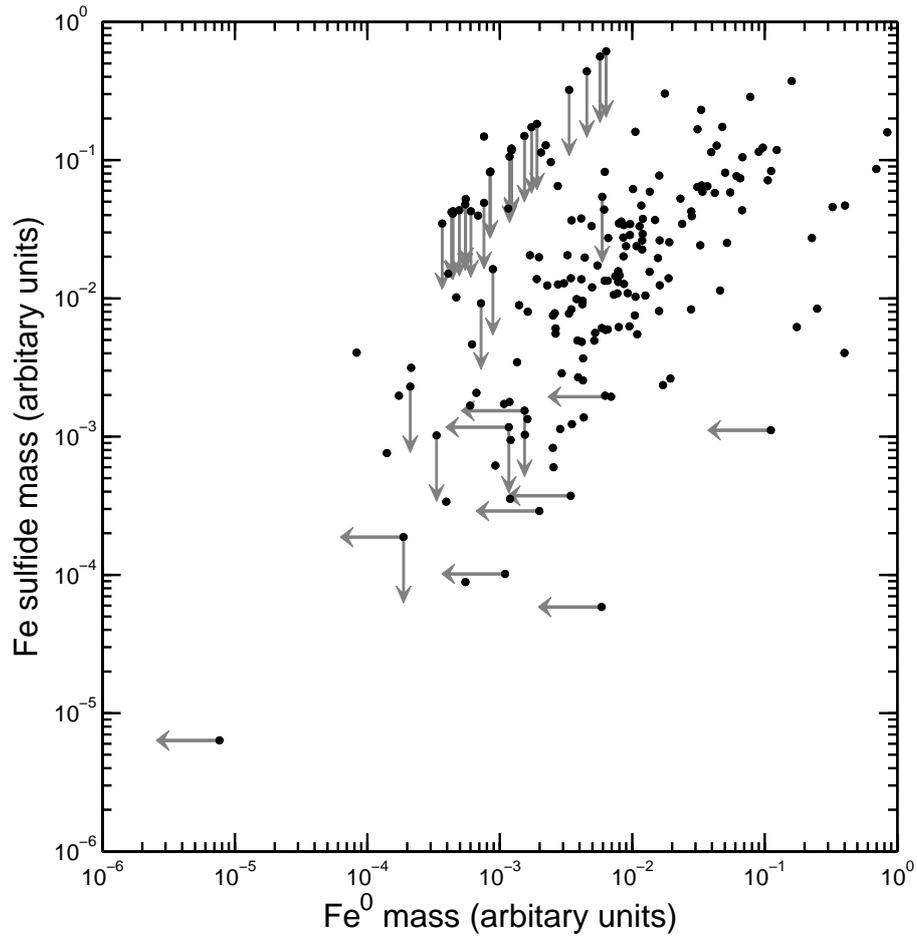}
        \caption{Fe sulfide mass vs. Fe$^{0}$ mass in the 194 Stardust fragments. We assume a detection limit of 1\% for components of a particular Fe oxidation state--the content of fragments with Fe$^{0}$ or Fe sulfide below the detection limit is given by a gray arrow.}
\label{figure:FeCorrelation}
  \end{figure}

The fragments we analyzed for this study were the larger Fe-bearing fragments in each track. Because Fe XANES analysis requires a strong Fe signal, we did not analyze the smaller particles near the bulb of the track which likely experienced more thermal alteration during the capture process. We preferentially analyzed larger particles in the bulb and the terminal particles which likely retained the oxidation state they had in the comet.  If Fe sulfides transformed to Fe metal upon aerogel capture, we should see an anticorrelation between the total amount of Fe sulfide in a Stardust fragment and the total amount of Fe$^{0}$ in the fragment: larger Fe sulfides would have a small fraction of their Fe converted to Fe$^{0}$ due to capture heating, smaller Fe sulfides, which can be severely thermally altered, would have a large fraction of their Fe converted from Fe sulfide to Fe$^{0}$. Multiplying the absorption edge jumps at the Fe K-edge (approximately proportional to Fe mass, as described earlier) by the fraction of Fe in Fe$^{0}$ and Fe sulfide, we can see if this effect is conspicuous in our data. Figure \ref{figure:FeCorrelation} shows a possible weak correlation, and certainly no anticorrelation, present in our data. We conclude that our measurements are not strongly affected by the reduction of Fe sulfides to Fe$^{0}$ as described by \citet{Ishii:2008p962}. The fragments we analyzed are probably large enough that this effect is not significant. For this reason, the reduction of oxidized Fe to Fe metal \citep{Leroux:2009p3050} is also unlikely to be a significant source of error in our determination of the bulk Fe oxidation state of comet Wild 2.

Our initial analysis of the IDPs included one particle, L2021v1cl6, that appeared more similar to chondritic-smooth (CS) than to chondritic-porous as shown in \citet{Brownlee:1985p2339}. The Fe XANES of this particle fits to mostly ferrihydrite, a hydrated Fe mineral. All of the other 15 IDPs have a more porous texture and Fe K-edge XANES spectra that fit to much smaller amounts of hydrated Fe minerals. We therefore exclude the IDP L2021v1cl6 from our data and conclude that the other 15 CP-IDPs are likely anhydrous.
 
\section{Discussion}
The presence of Fe in its different oxidation states is one of the clearest mineralogic indicators of the oxidation state of meteoritic material \citep{Rubin:1988p2665}. As discussed earlier, the Wild 2 samples and CP-IDPs are unlikely to have experienced much parent-body processing that would drastically change their oxidation state, and therefore a measurement of the Fe oxidation state of these two materials is a probe into the oxygen fugacity of the environment in which they formed. Figure \ref{figure:ternary} shows that comet Wild 2 and CP-IDPs differ in their Fe oxidation state at more than the 2$\sigma$ level. 

\citet{Ishii:2008p962} reported several differences between CP-IDPs and the Wild 2 material: the Stardust material lacks GEMS (glass with embedded metal and sulfides) and also lacks elongated enstatite whiskers and platelets, all of which are present in CP-IDPs. They conclude from these transmission electron microscopy observations that the Wild 2 material is dissimilar to CP-IDPs and more closely resembles asteroidal material. Our results, analyzing $\sim$1000$\times$ more material, indicate that comet Wild 2 differs at more than the 2$\sigma$ confidence level from CP-IDPs, and Wild 2 is also distinct from meteorites. CP-IDPs are 2$\sigma$ inconsistent with all meteorite groups except CI and R.

The significant difference between the Fe oxidation state of Wild 2 and CP-IDPs is that a significant fraction of Fe in Wild 2 exists as Fe$^{0}$ whereas CP-IDPs contain little Fe$^{0}$ compared to Fe sulfide and oxidized Fe. We can compare these results only qualitatively to, for example, TEM observations of mineral phases present in CP-IDPs: the Fe XANES technique used in this work can determine how many Fe atoms in the entire CP-IDP are in a given Fe oxidation state, whereas the the entire volume of the IDP is not always examined by TEM, and the amount of material in a given phase is seldom recorded. \citet{MacKinnon:1987p2660} tabulate a collection of phases observed in 30 CP-IDPs. Of the 30, Fe-bearing mineral phases are reported in 23 of the CP-IDPs: 21 contain oxidized Fe, 15 contain Fe sulfide, and 4 contain Fe$^{0}$. Qualitatively, this is consistent with the relative paucity of Fe$^{0}$ we see by the Fe XANES technique employed in this work. 

\subsection{Selection bias in CP-IDP collection?}

Since Fe metal is denser than Fe sulfides or oxidized Fe, particles containing substantial amounts of Fe metal may have a shorter residence time in the stratosphere and are therefore less likely to be collected. Particles containing Fe metal are also more likely to vaporize than less-dense particles \citep{Flynn:2008p3176}. 

Can the effect of these biases skew the composition of collected IDPs such that our observations of the 15 IDPs studied in this work are actually consistent with IDPs of Wild 2 composition before atmospheric entry? To answer this question, we recalculate the Fe oxidation for the 11 Stardust tracks from comet Wild 2 as if they were IDPs, accounting for the effects of atmospheric heating and residence time in the stratosphere.

First, we first calculate the residence time of 11 IDPs in the range of 15--25 km of altitude from the formulation given in \citet{Kasten:1968p3375} and \citet{Rietmeijer:1993p3303}. The IDPs have the Wild 2 Fe composition as given by the Stardust tracks in Table \ref{table:fefrac}. The speed of the particle falling through the atmosphere is proportional to its density, so the less-dense particles are more likely to be collected than the more-dense, Fe-metal-rich particles. Secondly, we consider the atmospheric entry heating of IDPs as described by \citet{Fraundorf:1980p3177}. We calculate the probability that an IDP is destroyed by heating to a temperature above the melting point of Fe metal, 1811$^{\circ}$ K.

The probability of collecting IDP $i$ of all $j$ types of IDPs can be expressed simply as:

\begin{equation}
P_i=\frac{\left(N_i-N_iV_i\right)T_i}{\displaystyle\sum_j{\left(N_j-N_jV_j\right)T_j}}
\label{entryeqn}
\end{equation}

where $N_i$ is the number of IDPs of type $i$, $V_i$ is the fraction of IDPs $i$ that vaporize, and $T_i$ is the residence time of IDP $i$ in the stratospheric-collection region.

For each of the 11 IDPs of Wild 2 composition, we calculate the probability of collection. Then we calculate the mean and 2$\sigma$ confidence limits of the Fe oxidation state of these 11 particles using the method described in Section \ref{confidencelimits}, but here each IDP is weighted by its probability of collection relative to the others.

The amount of Fe metal in IDPs of Wild 2 composition, taking into account atmospheric heating and residence time in the stratosphere, is $<$1\% less than the abundance of Fe metal in the IDPs before atmospheric entry. This cannot explain the observed lack of Fe metal in CP-IDPs compared to comet Wild 2, so we conclude that these possible selection biases in IDP collection are insignificant.

\subsection{Grain sizes in CP-IDPs and Stardust Fragments}
CP-IDPs are aggregates of mostly submicron grains whereas the Stardust fragments contain many mono-mineralic grains larger than a micron. Thermal modification or destruction of submicron grains in the bulb \citep{Leroux:2009p3050,Brownlee:2006p1376} precludes the possibility of comparing the Fe oxidation state of CP-IDPs and Stardust at the same grain size. It may be the case that cometary metal exists only in larger grains and is not present in the submicron components that make up CP-IDPs. While we cannot rule out this possibility with the measurements presented here, we can compare comet Wild 2 and CP-IDP material by looking at smaller grain sizes of Wild 2 (What is the Fe$^{0}$ fraction of the smaller Stardust fragments we analyzed?) and larger grain sizes of CP-IDPs (Do larger mineral grains of CP-IDPs exist and what is their Fe oxidation state?).

\begin{figure}
        \centering
        \includegraphics[width=\textwidth]{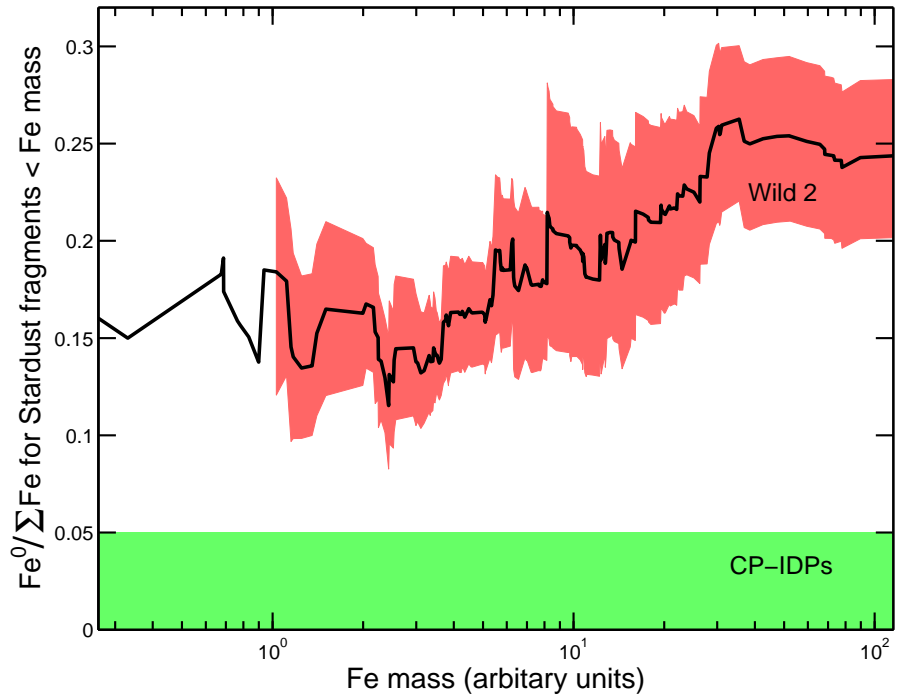}
        \caption{Fe$^{0}/\sum\,$Fe for all Stardust fragments less than a given Fe mass (computed by the Fe K-edge XANES jump). The 2$\sigma$ confidence interval for each set of Stardust fragments less than a given mass is shown by the red shaded region. The green region indicates the 2$\sigma$ confidence interval for the CP-IDPs measured in this work (CP-IDPs consist of mostly submicron grains--smaller grain size than the Stardust fragments analyzed in this work). Both the Stardust and CP-IDP uncertainties shown here are calculated with the particle-based approach described in Section \ref{confidencelimits}.}
\label{figure:FeMassFrac}
  \end{figure}

For each Stardust fragment we analyzed, we know its approximate, relative Fe mass by the absorption edge jump across the Fe K-edge. We compute Fe$^{0}/\sum\,$Fe for all Stardust fragments less than a given Fe mass, shown in Figure \ref{figure:FeMassFrac}. The Fe$^{0}$ fraction increases for the intermediate-mass particles in this study, but is still inconsistent with the CP-IDPs over the entire Fe-mass range of Stardust fragments we analyzed (using the particle-based bootstrap uncertainty calculation).

Stratospheric collection yields large ($>$5$\mu$m) mono-mineralic grains of non-chondritic composition in addition to IDPs. Fine-grained chondritic material (reminiscent of small bits of CP-IDPs) is sometimes found adhering to the surfaces of these grains, indicating they originate from the same parent body or bodies as CP-IDPs \citep{Flynn:2007p3428,Flynn:2008p3176}. The sizes of these grains are closer to the range of grain sizes of the Stardust fragments we analyzed. The mono-mineralic IDP grains consist predominantly of olivine, pyroxene, and sulfide. They probably do not contain a significant amount Fe metal. Therefore, these larger mono-mineralic IDP grains which likely sample the same parent body as the CP-IDPs, probably do not contain enough Fe$^{0}$ to be consistent with the comet Wild 2 material.

If a cosmic dust particle is highly enriched in Fe metal to the exclusion of most other phases, its chemical composition would not be close to chondritic and therefore the particle would not be considered a CP-IDP. Similarly, a particle made mostly of Fe metal may not appear fluffy, and so we may not select it from the Cosmic Dust Catalogs for inclusion in a study such as this. There is an inherent bias in the selection of CP-IDPs for comparison with other samples of matter. Stratospheric-collected particles with high Fe metal abundances have been classified as ``Cosmic", but the fraction of the collected Fe metal particles which are of true extraterrestrial origin or possibly cometary has not been established. However, if Fe metal were present as submicron grains at the 10\% level, a particle could still have approximately chondritic composition and appear fluffy, like a CP-IDP. This, we have shown, is not the case. For this type of selection bias to significantly affect our results, the Fe metal distribution in IDPs would have to be very heterogenous: the Fe metal would need to be sequestered into particles which contain enough Fe metal to appear either non-chondritic or non-porous.

Our analysis compares sub-micron-grained CP-IDPs with larger-grained Stardust fragments, but as the Stardust fragments show more Fe$^{0}$ than CP-IDPs over the entire range of particle sizes we measured, and large mono-mineralic grains which likely originated from the same source material as IDPs also show a dearth of Fe$^{0}$, we conclude it is likely that the CP-IDPs contain less Fe$^{0}$ than the Wild 2 samples at comparable grain sizes. We acknowledge a possible selection bias in choosing chondritic and porous IDPs for this study, though our results are valid under the assumption that Fe metal is not distributed heterogeneously, with scarce, very Fe-metal-rich particles containing the Fe metal missing in CP-IDPs.

\subsection{Fe metal in Stardust samples}
\begin{figure}
        \centering
        \includegraphics[width=\textwidth]{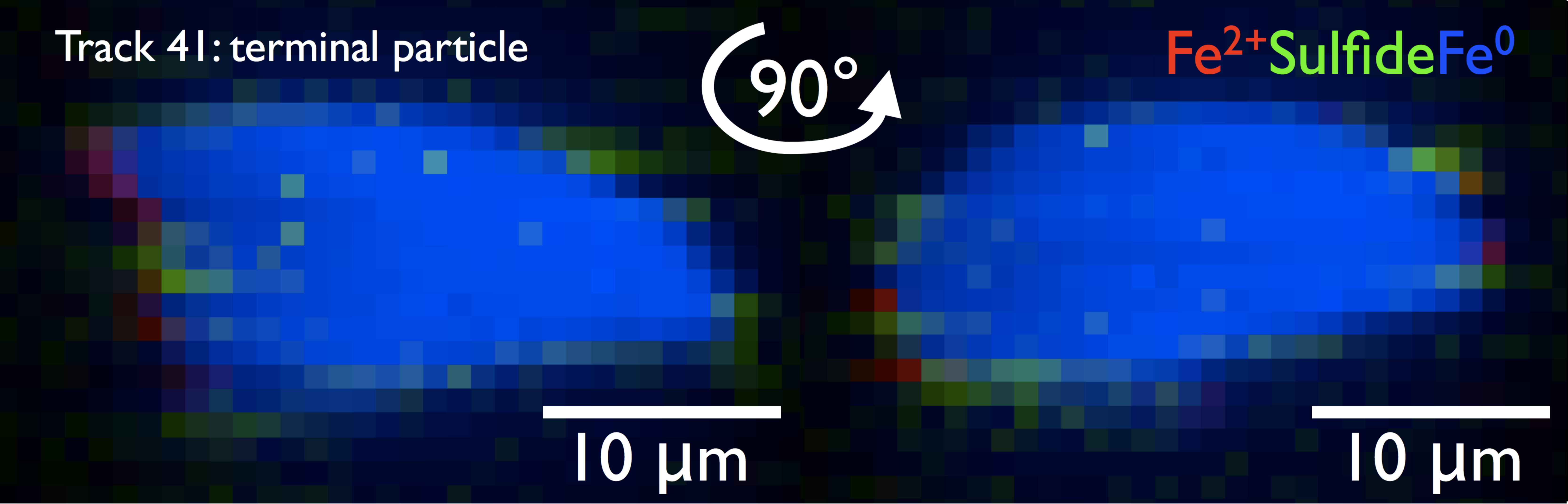}
        \caption{Iron chemical map \citep{Ogliore:2008p3135} of the the large terminal particle in Track 41, ``Simeio", in two orientations that differ by a 90$^{\circ}$ rotation. Simeio is mostly a large mass of Fe$^{0}$. The mineral phase of this material was identified as kamacite and taenite by x-ray diffraction (XRD). Iron sulfide is present in small amounts throughout the particle, as well as both Fe$^{2+}$ and sulfide on the particle perimeter in both orientations, possibly indicating a thin, partial shell of sulfide and Fe$^{2+}$. }
\label{figure:simeio}
  \end{figure}

The Fe metal we observe in the Wild 2 samples is unlikely to be a capture effect, as discussed earlier. \citet{Zolensky:2006p1378} report that the high abundance of Ni in metal droplets in Stardust tracks shows that there must have been metal intrinsic to the comet. Much of the Fe metal we measure exists in large particles in the terminus of the track. One of the terminal particles in Track 77 was observed in TEM to be an Fe metal grain encased in olivine (D. Joswiak, private communication), reminiscent of Fe metal encapsulated in silicate grains in unequilibrated chondrites \citep{Rambaldi:1980p2671}. The terminal particle of Track 41 is a large piece of Fe metal (as shown by the Fe XANES, and an x-ray diffraction match to kamacite and taenite) with a bit of Fe sulfide and Fe$^{2+}$ (as seen in the chemical map of the particle, Figure \ref{figure:simeio}). The Fe metal in large terminal particles like these almost certainly existed as metal in the comet, which makes Wild 2 distinct from CP-IDPs. 

\subsection{Nebular mixing, CP-IDPs, and comet Wild 2}
The oxidation state of Fe can be used as an indicator of the oxygen fugacity in which the grain formed \citep{Rubin:1988p2665}. These metal grains likely formed in the high-temperature, low-O-fugacity conditions of the inner region of the early solar nebula (a reducing environment \citep{Wooden:2008p2674}). Refractory inclusions from the inner nebula have been identified in the Stardust samples \citep{Simon:2008p2065}. Our Fe XANES survey of Stardust tracks imply that a fraction between 0.5 and 0.65 (2$\sigma$) of Wild 2 material came from the inner nebula \citep{Westphal:2009p1874}, suggesting that there was large-scale mixing in the solar nebula \citep{Ogliore:2009p2695}. Stardust olivines show a broad distribution of Mg/Fe composition \citep{Zolensky:2006p1378} which indicates the comet consists of grains that formed in a wide range of reducing/oxidizing conditions \citep{Rubin:1988p2665}. Therefore, it is not unexpected that the oxidation state of Fe in Wild 2 should reflect both the inner nebula (Fe metal grains) and the outer nebula (Fe-rich silicates \citep{Wooden:2008p2674}).  

The fact that CP-IDPs show much less Fe metal than Wild 2 indicate that these particles did not originate from a body, or bodies, that sampled the regions of low oxygen fugacity that Wild 2 did. Comets can contain different amounts of low and high oxygen fugacity material, as \citet{Wooden:2008p2674} states: ``Hence, the efficiency of incorporation of low vs. high oxygen fugacity products into cometary nuclei depends on the interplay between the duration of low vs. high water vapor content, the time-dependent outward mass transport rate, and the distance(s), range of radial migration, and duration of cometesimal accretion and nuclei accumulation". CP-IDPs could be derived from comets that did not sample the highly mixed nebula as Wild 2 did, perhaps one that formed in a different time or place.

Our results are inconsistent with the hypothesis that CP-IDPs originate from parent bodies with the composition of comet Wild 2 at $>$2$\sigma$ confidence level. We have not, however, disproved the hypothesis that CP-IDPs originate from Jupiter-family comets because we do not know if all Jupiter-family comets have an Fe oxidation state that is consistent with comet Wild 2.

Comet Tempel 1, the target of NASA's Deep Impact mission, shows significant differences from Wild 2 in composition \citep{Brownlee:2006p1376} and morphology of the cometary nucleus \citep{Thomas:2007p3161}. These differences could be due to the longer period of time Tempel 1 spent in the inner solar system, the different regions from which the samples originated, the different size-range of analyzed material, and many other complicating factors. However, it is also possible that the two comets are composed of fundamentally different material that formed in different places and at different times in the solar nebula. From orbital dynamics, \citet{Liou:1996p3165} show that dust from Tempel 1 could contribute significantly to the population of low-entry-speed CP-IDPs collected at Earth. This work has shown that a cometary source of CP-IDPs must contain less inner nebula material (where the oxygen fugacity is low) than comet Wild 2--we do not rule out a CP-IDP source from comet Tempel 1 or the numerous other comets of unknown composition.

\section{Acknowledgements}
The authors thank J. Warren and R. Bastien at Johnson Space Center for skillful extraction and mounting of the IDPs, and the reviewers for helpful comments. We thank M. Humayun for permission to publish the characterization of Simeio. This work was supported by NASA SPS and DDAP grants. We gratefully acknowledge the provision of XANES spectra of standard compounds by D. L. Bond, T. Borch, B. C. Bostick, T. Buonassisi, M. Burchell, C. S. Chan, S. Fendorf, C. Hansel, M. Heuer, A. Kearsley, C. S. Kim, M. Newville, P. Nico, P. O\textquoteright Day, N. Rivera, K. Ross, C. Santelli, B. Toner, and G. A. Waychunas. The Advanced Light Source is supported by the Director, Office of Science, Office of Basic Energy Sciences, of the U.S. Department of Energy under Contract No. DE-AC02-05CH11231.

\singlespacing

\begin{sidewaystable}[ht]
\centering
\begin{tabular}{l c c c c c c}
\hline\hline  \TT \BB
Particle or Track & \parbox[c]{5pc}{Fragments Analyzed} & \parbox[c]{5pc}{Fraction Fe$^0$} & \parbox[c]{5pc}{Fraction Sulfide} &  \parbox[c]{5pc}{Fraction Oxidized} &   \parbox[c]{5pc}{Fraction Magnetite}  &  \parbox[c]{4pc}{Relative Fe Mass} \\  [0.5ex]
\hline
\T IDP L2006v1cl14 & 1 & 0.00 & 0.75 & 0.25 & 0.04 & 0.37 \\
IDP L2008ad1cl17 & 1 & 0.01 & 0.00 & 0.99 & 0.00 & 0.16 \\
IDP L2008ae1cl5 & 1 & 0.00 & 0.48 & 0.52 & 0.00 & 0.63 \\
IDP L2011a11cl8 & 1 & 0.07 & 0.25 & 0.69 & 0.00 & 0.11 \\
IDP W7066d1cl1 & 1 & 0.00 & 0.93 & 0.07 & 0.01 & $\equiv$1 \\
IDP L2036aq1cl17 & 1 & 0.00 & 0.15 & 0.85 & 0.00 & 0.32 \\
IDP L2036ap1cl19 & 1 & 0.04 & 0.00 & 0.97 & 0.00 & 0.14 \\
IDP L2036AF2217 & 1 & 0.00 & 0.50 & 0.50 & 0.04 & 2.8 $\times$ 10$^{-2}$ \\
IDP U2015Hz52 & 1 & 0.12 & 0.18 & 0.70 & 0.00 & 1.1 $\times$ 10$^{-2}$ \\
IDP L2036AF3208 & 1 & 0.01 & 0.27 & 0.72 & 0.00 & 5.0 $\times$ 10$^{-2}$ \\
IDP L2036AF4199 & 1 & 0.03 & 0.11 & 0.87 & 0.00 & 1.5 $\times$ 10$^{-2}$ \\
IDP U2-20-GCA-7-2005 & 1 & 0.07 & 0.00 & 0.93 & 0.20 & 0.12 \\
IDP W7110A-2E & 1 & 0.01 & 0.76 & 0.24 & 0.00 & 0.14 \\
IDP W7031f1cl1 & 1 & 0.05 & 0.01 & 0.94 & 0.00 & 2.4 $\times$ 10$^{-2}$ \\
IDP W702901cl2 & 1 & 0.00 & 0.00 & 1.00 & 0.00 & 4.6 $\times$ 10$^{-3}$ \\ [0.5ex]
\hdashline
\T Stardust Track 38 & 3 & 0.18 & 0.05 & 0.75 & 0.00 & 3.4 $\times$ 10$^{-2}$ \\
Stardust Track 41 & 33 & 0.30 & 0.52 & 0.18 & 0.00 & $\equiv$1 \\
Stardust Track 42 & 3 & 0.00 & 0.96 & 0.04 & 0.00 & 4.1 $\times$ 10$^{-5}$ \\
Stardust Track 43 & 4 & 0.16 & 0.29 & 0.54 & 0.00 & 4.7 $\times$ 10$^{-2}$ \\
Stardust Track 44 & 1 & 0.89 & 0.10 & 0.01 & 0.00 & 6.8 $\times$ 10$^{-4}$ \\
Stardust Track 54 & 2 & 0.10 & 0.09 & 0.81 & 0.00 & 5.0 $\times$ 10$^{-4}$ \\
Stardust Track 74 & 56 & 0.10 & 0.38 & 0.52 & 0.00 & 0.29 \\
Stardust Track 77 & 53 & 0.16 & 0.36 & 0.48 & 0.00 & 8.2 $\times$ 10$^{-2}$ \\
Stardust Track 103 & 2 & 0.55 & 0.28 & 0.17 & 0.00 & 1.6 $\times$ 10$^{-5}$ \\
Stardust Track 104 & 6 & 0.05 & 0.28 & 0.66 & 0.00 & 2.2 $\times$ 10$^{-4}$ \\
Stardust Track 105 & 31 & 0.11 & 0.61 & 0.28 & 0.00 & 4.3 $\times$ 10$^{-2}$ \\
\hline
\end{tabular}
\caption{Fraction of Fe in Fe$^0$, sulfide, and oxidized Fe, for the IDPs and Stardust tracks analyzed. Also listed is the relative Fe mass in each IDP and Stardust track, and the fraction of magnetite in each particle or track. For the Stardust tracks we assume smelting is negligible, as described in the text.}
\label{table:fefrac}
\end{sidewaystable}

\doublespacing

\section{References}
\bibliographystyle{elsarticle-harv}
\bibliography{papers_bibtex}

\end{document}